\documentclass[12pt]{iopart}

\usepackage{graphicx}
\usepackage{dcolumn}

\begin{document}

\title[Control parameters]{Control parameters in turbulence, Self Organized Criticality and ecosystems.}

\author{S C Chapman\dag, G Rowlands\dag and N W Watkins$\sharp$}
\address{\dag Centre for Fusion, Space and Astrophysics, Physics Dept., University of Warwick, UK}

\address{$\sharp$ British Antarctic Survey (NERC), Cambridge, UK}

\date{\today}
\ead{S.C.Chapman@warwick.ac.uk}


\begin{abstract}
    From the starting point of the well known Reynolds number of fluid turbulence we propose a control parameter $R$ for a wider class of systems including avalanche models that show Self Organized Criticality (SOC) and ecosystems. $R$ is related to the driving and dissipation rates and from similarity analysis we obtain a relationship $R\sim N^{\beta_N}$ where $N$ is the number of degrees of freedom. The value of the exponent $\beta_N$ is determined by detailed phenomenology but its sign follows from our similarity analysis. For SOC, $R=h/\epsilon$ and we show that $\beta_N<0$ hence we show independent of the details that the transition to SOC is when $R \rightarrow 0$, in contrast to fluid turbulence, formalizing the relationship between turbulence (since $\beta_N >0$, $R \rightarrow \infty$) and SOC ($R=h/\epsilon\rightarrow 0$).  A corollary is that SOC phenomenology, that is, power law scaling of avalanches, can persist for finite $R$ with unchanged exponent if the system supports a sufficiently large range of lengthscales; necessary for SOC to be a candidate for physical systems.  We propose a conceptual model ecosystem where $R$ is an observable parameter which depends on the rate of throughput of biomass or energy; we show this has $\beta_N>0$, so that increasing $R$ increases the abundance of species, pointing to a critical value  for species 'explosion'.
\end{abstract}

\maketitle

\section{Introduction}
A central idea in physics is that complex, often intractable, behavior can be
quantified by a few measurable control parameters. In fluid turbulence a single control parameter, the
Reynolds number $R_E$  quantifies the
transition from ordered (laminar) to disordered (turbulent) flow.
This parameter is from dimensional arguments a function of
macroscopic system variables, but is also expressible as a function of the number of energy carrying modes or
degrees of freedom (d.o.f.)\cite{frisch}. Although the detailed dynamics are different, many of the macroscopic
features of idealized, finite (large) Reynolds number turbulence are also characteristic of
other strongly correlated, out of equilibrium systems.
These systems  can
all be driven into a disordered state with defining characteristics: they
have many degrees of freedom (d.o.f.); are driven and dissipating, are
out of equilibrium but on average in a steady state, and  show anomalous
scaling over a large dynamic range. Loosely speaking, a `class' of such systems
will show an order- disorder transition, captured by varying a single control parameter.
This includes avalanche models exhibiting Self Organized Criticality
(SOC)\cite{BTW87,frette,jensenbook,sornette,sethna,dickman}.
It was originally
argued\cite{BTW87,BTW88} that these systems self organize to an
SOC state which in the above sense is
their state of maximal disorder. Subsequent analysis has
established a consensus\cite{vesp,vesp2,vergeles,sethna,sornette} that
SOC is a limiting behavior in the driving rate $h$ and the
dissipation rate $\epsilon$, such that $h/\epsilon \rightarrow 0$ with
$h$,$ \epsilon \rightarrow 0$, (and $h \leq \epsilon$, that is,
steady state).
  This is exemplified by the constructive
  definition (in\cite{jensenbook}) as ``slowly driven interaction dominated thresholded"
  (SDIDT) systems.

 In this paper we give a prescription for obtaining a control parameter for these systems, in analogy to the Reynolds number in fluid turbulence.
For avalanche models exhibiting SOC, we identify d.o.f. with realizable avalanche sizes and we show that the relevant control parameter $R_A$ is
$h/\epsilon$. It follows that the SDIDT
 limit is reached by taking
$R_A$ to zero; we show this maximizes the number of d.o.f. in the
opposite sense to fluid turbulence. This result clarifies the much
debated relationship between turbulence and
SOC\cite{turbsoc,bofetta,bramnature,bramprl,bramN,chapjpa}. An important
corollary is that SOC phenomenology can quite generally persist
under conditions of finite drive in a sufficiently large bandwidth
system.
As our result flows from dimensional analysis it is quite generic.
An important example that we give here is as a parametrization of ecosystem models for species
abundance\cite{ecobook1,ecobook2,allenscience}.
For ecosystems we show that as the control parameter increases so does the
abundance of species, or d.o.f. This
points to the possibility of a critical value at which
 the onset of
diversification of species occurs.

\section{Similarity analysis and Reynolds number}
The systems that we have in mind all have strongly coupled d.o.f. that
transport some quantity from the driving to the
dissipation scale. Provided that this 'dynamical quantity' is
governed by a conservation law to ensure  steady state (not necessarily equilibrium) on the
average we insist that its precise nature, and the microscopic details of how it
is transported are not relevant to the macroscopic ensemble
average behavior. We seek a control parameter expressible in terms of the
number of d.o.f. of the system that parameterizes the transition from ordered (few d.o.f.) to disordered
(many d.o.f.) behavior.
We identify the control parameters of the
system in terms of known macroscopic variables by  formal dimensional analysis
(similarity analysis or Buckingham $\Pi$ theorem, see e.g.
\cite{barenblatt}). Any
system's behavior is captured by a general function $F$ which
only depends on the \emph{relevant} variables $Q_{1..V}$ that
describe the system. Since $F$ is dimensionless it must be a function of the possible dimensionless groupings
$\Pi_{1..M}(Q_{1..V})$ which can be formed from the $Q_{1..V}$.
The (unknown) function $F(\Pi_1,\Pi_2,..\Pi_M)$ is universal,
describing all systems that depend on the $Q_{1..V}$ through the
$\Pi_{1..M}(Q_{1..V})$ and the relationships between them. If one
then has additional information about the system, such as a conserved
quantity, the $\Pi_{1..M}(Q_{1..V})$  can be related to each other
to make $F$ explicit. Thus this method can lead to information
about the solution of a class of systems where the governing
equations are unavailable or intractable, often the case
for complex systems where there are a large number ($N$
here) of strongly coupled d.o.f.. If the $V$ variables are
expressed in $W$ dimensions (i.e. mass, length, time) then there
are $M=V-W$ dimensionless groupings.

Here, since we have that
the precise nature of the transported dynamical quantity is
irrelevant, the only relevant dimensions are length and time so
that $W=2$. We next insist that there is a single control
parameter ($R$, in the case of turbulence, the Reynolds number $R_E$) which may be
expressed as a function of the number of active degrees of freedom
$N$. This means that the system's behaviour is captured by some
$F(\Pi_1,\Pi_2)$; where $R=\Pi_1$ and $\Pi_2=f(N)$ and the
$\Pi_1$ and $\Pi_2$ are related via some conservation property.
Hence $M=2$ so that $V=4$; there are four relevant variables to
consider.

It is useful to fix ideas in terms of a relatively well understood example of the above, namely turbulence. Our aim here is to obtain a control parameter $R$
by analogy to $R_E$ via dimensional analysis; for a detailed discussion of the universal scaling properties of Kolmogorov (K-41) turbulence and their origin in the Navier Stokes equations see for example \cite{frisch}. As above, for K-41  we have four relevant
macroscopic variables (given in Table 1) and
two dimensionless groups:
\begin{equation}
\Pi_1=\frac{UL_0}{\nu}=R_E,\hskip 2pt
\Pi_2=\frac{L_0}{\eta}=f(N)
\end{equation}
\begin{table}[t]
\begin{center}
\caption{$\Pi$ theorem applied to homogeneous
turbulence.}
\begin{tabular}{ccl}
\hline
Variable      &dimension    & description  \\
\hline
$L_0$&$L$& driving length scale\\
$\eta$&$L$&dissipation length scale\\
$U$&$LT^{-1}$&bulk (driving) flow speed\\
$\nu$&$L^2T^{-1}$&viscosity\\
\hline
\end{tabular}
\label{tab1}
\end{center}
\end{table}
$\Pi_1$ is the Reynolds number of the flow, and the ratio of lengthscales
$\Pi_2$ is directly  related to the number of d.o.f. $N$
available and we now relate $R_E$ to $f(N)$ (or $\Pi_1$ to $\Pi_2$). For
incompressible fluid turbulence, our dynamical quantity is the
time rate of energy transfer per unit mass $\varepsilon_l$ through
length scale $l$. Conservation
and steady state imply that in an ensemble averaged sense this is
balanced by the rate at which energy is transferred to the fluid
$\varepsilon_{inj}\sim U^3/L_0$ which in turn is balanced by the
dissipation of energy within the fluid $\varepsilon_{diss}$ so
that $\varepsilon_{inj}\sim\varepsilon_l\sim\varepsilon_{diss}$.
Dimensional arguments (e.g. \cite{frisch})  lead to
$\varepsilon_{diss}\sim \nu^3/\eta^4$ . Conservation, that is
$\varepsilon_{inj}\sim\varepsilon_{diss}$ then gives the well
known result \cite{frisch} which relates $\Pi_1$ and $\Pi_2$:
\begin{equation}
R_E=\frac{UL_0}{\nu}\sim\left(\frac{L_0}{\eta}\right)^\frac{4}{3}
\end{equation}
The $4/3$ exponent that arises for K-41 is modified if we consider other turbulent flows with different
phenomenologies, for example
 anisotropy, and intermittency. Nevertheless, for any turbulent flow we can anticipate that
the relationship between $L_0/\eta$ and the number of degrees of freedom $N$ will be of the form:
\begin{equation}
N \sim \left(\frac{L_0}{\eta}\right)^\alpha
\end{equation}
with $\alpha >0$; the crucial point is that for turbulence, $N$ always
grows with $L_0/\eta$. The only property of turbulence with which we are concerned here is that
\begin{equation}
R_E\sim \left(\frac{L_0}{\eta}\right)^\beta \sim N^{\beta_N}
\end{equation}
and that in particular, for turbulence $\beta_N=\beta \alpha >0$.
This identifies the  Reynolds number as the
control parameter for a process (turbulence) which
simply grows more active modes or d.o.f. as we increase $R_E$,
 taking the system from order
(few d.o.f. or laminar flow) to disorder (many coupled d.o.f.).

We will now see that more generally, similarity analysis is sufficient to obtain the relationship between the control parameter $R$ and
the number of degrees of freedom $N$ of the form:
\begin{equation}
R \sim N^{\beta_N}
\end{equation}
 The value of the exponent $\beta_N$ will depend on the details of these systems but crucially we will see that
the sign of $\beta_N$ is obtained from similarity analysis. This is sufficient to establish if, as in the case of turbulence, increasing $R$ increases the
disorder or complexity of the system.

\section{Control parameter for avalanching systems}

 We now envisage a generic avalanche model in a system of size
$L_0$ where the height of sand is specified on a grid, with nodes
at spacing $\delta l$. Sand is added to individual nodes, that is,
on length scale $\delta l$  at an average time rate
$\varepsilon_{inj}=h$ per node. There is some process, here
avalanches, which then transports this dynamical quantity (the
sand) though structures on intermediate length scales $\delta l
<l<L_0$. Sand is then
  lost to the system (dissipated) at a time rate $\epsilon$ over the
  system size $L_0$.
The relevant variables for the avalanching system are given in
Table 2. The two dimensionless
groups are:
\begin{equation}
\Pi_1=\frac{h}{\epsilon}=R_A,\hskip 2pt
\Pi_2=\frac{L_0}{\delta l}=f(N)
\end{equation}
The second parameter,
$\Pi_2=f(N)$ is related to the number of d.o.f. of the
system.
The control parameter $\Pi_1=h/\epsilon$ is analogous to the Reynolds number above in that, as we will now show,
it relates the ratio of the driving to the dissipation rates to the number of active, or energy containing degrees of freedom $N$ in the system.

 In Euclidean dimension $D$ there are $(L_0/\delta l)^D$
nodes so that
conservation\cite{decarvalho,decar2,decar3} of
the flux of sand (in an ensemble averaged sense), gives
$h(L_0/ \delta l)^D \sim \epsilon$
which simply states that the rate at which sand is added to the
system must on average balance the rate at which sand leaves.
\begin{table}[t]
\begin{center}
\caption{$\Pi$ theorem applied to an avalanching
system. The sand carries a property  with
dimension $S$.}
\begin{tabular}{ccl}
\hline
Variable        &dimension    & description  \\
\hline
$L_0$&$L$&system size\\
$\delta l$&$L$&grid size\\
$\epsilon$&$ST^{-1}$& system average dissipation/loss rate\\
$h$&$ST^{-1}$&average driving rate per node\\
\hline
\end{tabular}
\label{tab1}
\end{center}
\end{table}
On  intermediate length scales $\delta l <l<L_0$, sand is
transported via avalanches. There must be some detail of the
internal evolution of the pile that maximizes the number of length
scales $l$ on which avalanches occur. For avalanche models this is
the property that transport can only occur locally if some local
critical gradient is exceeded; as a consequence the pile evolves
through many metastable states.  If these length scales represent
d.o.f. then  the number $N$ of
d.o.f. available  will be bounded by $L_0$ and $\delta l$ so that:
\begin{equation}
N\sim (L_0/\delta l)^\alpha
\end{equation}
with $D \geq \alpha\geq 0$ for
$D>1$ ($\alpha$ may be fractional). We then have:
\begin{equation}
R_A=\frac{h}{\epsilon}\sim \left(\frac{\delta l}{L_0}\right)^D\sim
N^{-\alpha D}
\end{equation}
 This is in contrast to fluid turbulence since the number of d.o.f.
\emph{decreases} with increasing drive, that is, increasing
$R_A=h/\epsilon$. Thus we recover the SDIDT limit for SOC, namely
$R_A \rightarrow 0$, but now explicitly identify this limit with
maximizing the number of d.o.f. available, that is, the disorder
of the system. Our result from dimensional analysis
is to obtain $R_A \sim N^{\beta_N}$ and to show quite generally that $\beta_N<0$.
Following the above discussion of turbulence, we can go further and make the analogy $R_A \equiv R_E$, that is,
the system's 'effective Reynolds number' increases with the energy/sand taken up by the system, i.e. with $h$.

The property that the system generates many coupled d.o.f. is, for
SOC, captured by avalanching phenomenology. This sets conditions
on the microscopic details of the system; specifically, there must
be a separation of timescales in that the relaxation time for the
avalanches must be short compared to  the time taken for the
drive to on average cause a cell to be come unstable so that
avalanching is the dominant mode of transport. The critical
gradient can be a random variable but provided it has a defined
average value $g$, we have an average number of timesteps to drive
a cell unstable $(g \delta l)/(h \delta t)$ where $\delta t$ is
the timestep. This gives two conditions for
avalanching to dominate transport\cite{nickgrl}:
\begin{equation}
h \delta t \ll g \delta l,\hskip 2pt
h \delta t \ll g \delta l \left(\frac{L_0}{\delta l}\right)^D
\end{equation}
  The first
condition is that avalanches only occur after many grains of sand
have been added to any given cell in the pile and is the strict
SDIDT\cite{vesp,vesp2,vergeles}  limit. However, if the system has large
bandwidth $L_0\gg\delta l$,  one can consider an intermediate
behavior $g L_0 \gg h \delta t > g \delta l$ where the driver is large
enough to swamp the smallest avalanches, but larger avalanches
persist\cite{nickgrl}.  For fixed
$L_0$ and $\delta l$, increasing $h \delta t$ above $ g \delta l$
successively erodes the available d.o.f since each addition of
sand swamps $h \delta t/(g \delta l)$ cells of the pile. Ultimately as $h$
and hence $R_A$ is increased to the point where $h \delta t \sim g \delta l \left(L_0/\delta l\right)^D$  there will be a crossover to laminar flow.

We now show that this intermediate, finite $R_A$ behavior will be
 `SOC like', with power law avalanche statistics sharing the same exponent as at the SDIDT limit; confirming our assumption above that $\beta_N$ is independent of
 the control parameter $R_A$.
To see this, consider passing
through the regime of   $h \delta t \sim g \delta l$ with $h \delta t\ll g \delta l
\left(L_0/\delta l\right)^D$, which can be achieved by increasing
both $h$ and $L_0$ such that $h\rightarrow A h$ and
$L_0\rightarrow L_0A^{(1/D)}$. This is equivalent to coarse
graining the pile spatially, so that provided the system has self
similar spatial
 scaling  we can anticipate
obtaining the same solution (subject to a rescaling) provided  $L_0\rightarrow
L_0A^{(1/D)}$. Under nonlocal feeding and non overlapping avalanches ($A$ times as many grains added at well separated positions over the
pile) this coarse-graining may not occur.

 We illustrate this in
Figures 1 and 2 with simulations of the BTW\cite{BTW87} sandpile in 2D, where the
driving occurs randomly in time and is spatially restricted to the
`top' of the pile. In all cases the critical gradient (threshold for avalanching) is $g=4$.
Figure 1 shows two simulations of size $L_0=100$ with $h=[4,16]$ ($\delta t=1$ in the simulations) i.e. just at, and above, the regime $h \delta t \sim g \delta l$, but in both cases
with $h \delta t \ll g \delta l
\left(L_0/\delta l\right)^D$. At $h=16$ we see that the power law statistics of the smallest avalanches is lost but there is still scaling over a more restricted range of avalanche sizes, i.e. we have the same scaling, and same exponent, but over a reduced number of d.o.f.
Rescaling the avalanche sizes of the $h=16$ run $S \rightarrow S/16$
(that is, lengthscales $l\rightarrow l/4$) recovers the behaviour of the $h=4$ run except at the largest decade. To recover the full range we repeat the $h=16$ run in a larger box, $L_0=400$ (that is, $L_0 \rightarrow 4L_0$) which is shown alongside the $h=4$, $L_0=100$ run in Figure 2. Rescaling the $L_0=400$ run with $S \rightarrow S/16$, i.e.  lengthscales $l\rightarrow l/4$ then reproduces the $h=4$, $L_0=100$ results.
\begin{figure}
\includegraphics[width=0.4 \textwidth]{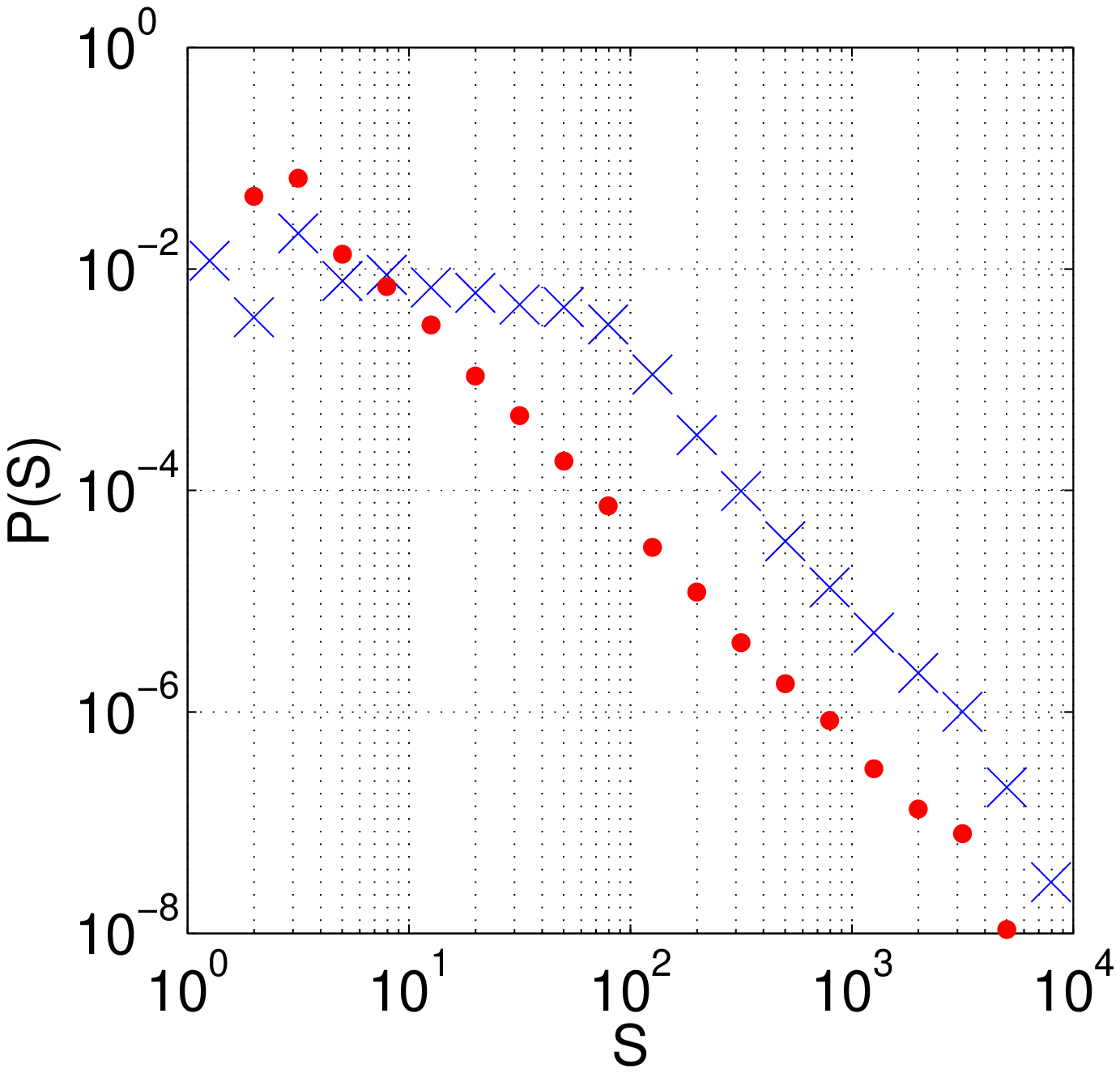}\hskip 3cm
\includegraphics[width=0.4 \textwidth]{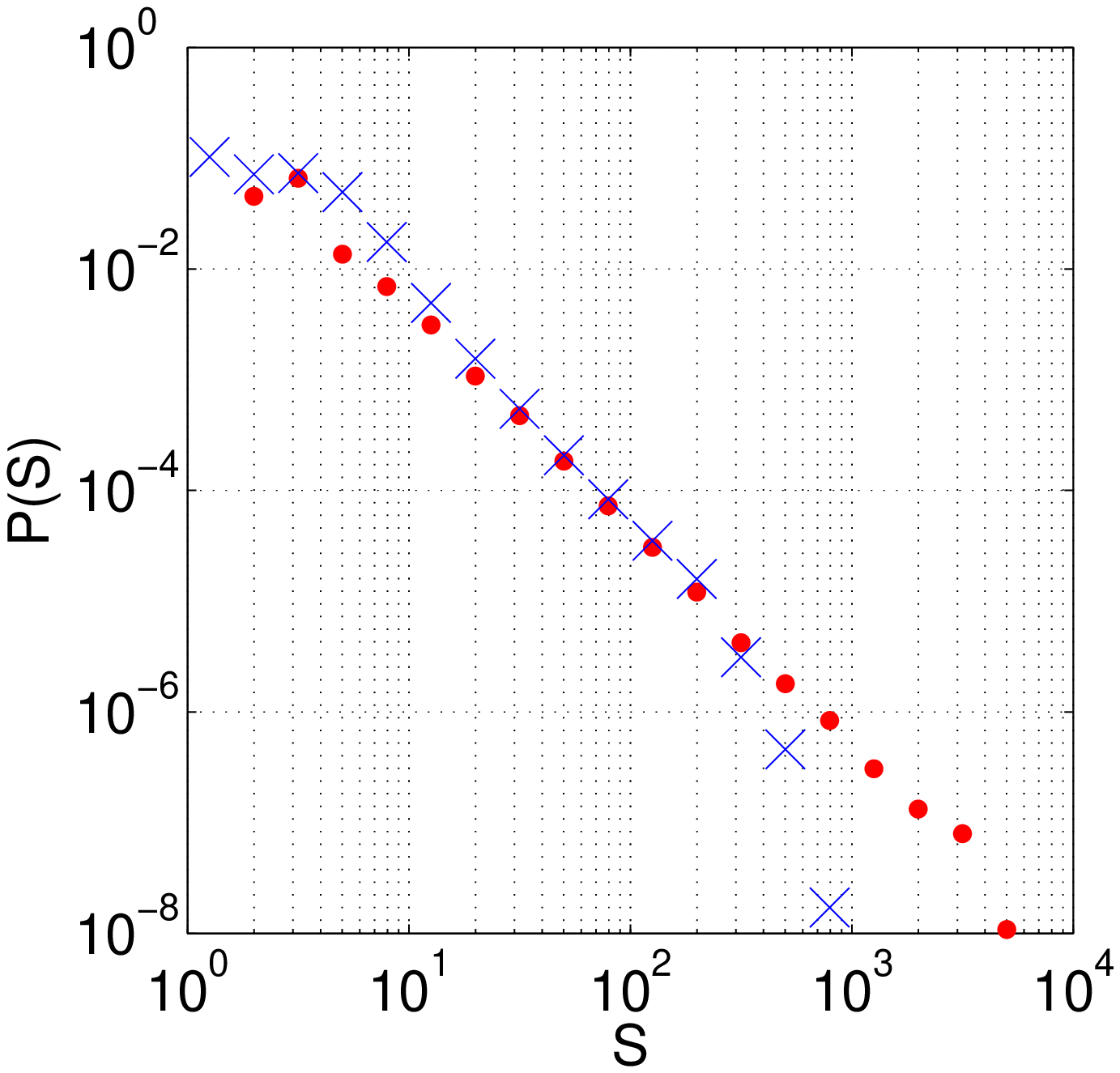}
\caption{Avalanche size normalized distributions for two runs of the 2D
BTW\cite{BTW87,BTW88} sandpile driven at the 'top' corner formed
by two adjacent closed boundaries, the other boundaries are open.
$L_0=100$ and $h=4$ ($\bullet$) and $h=16$ ($\times$); (a-left)
probability densities; (b-right) as (a) with probability density for the
$h=16$ avalanche sizes rescaled $S\rightarrow S/16$.}
\end{figure}
\begin{figure}
\centering
\includegraphics[width=0.4 \textwidth]{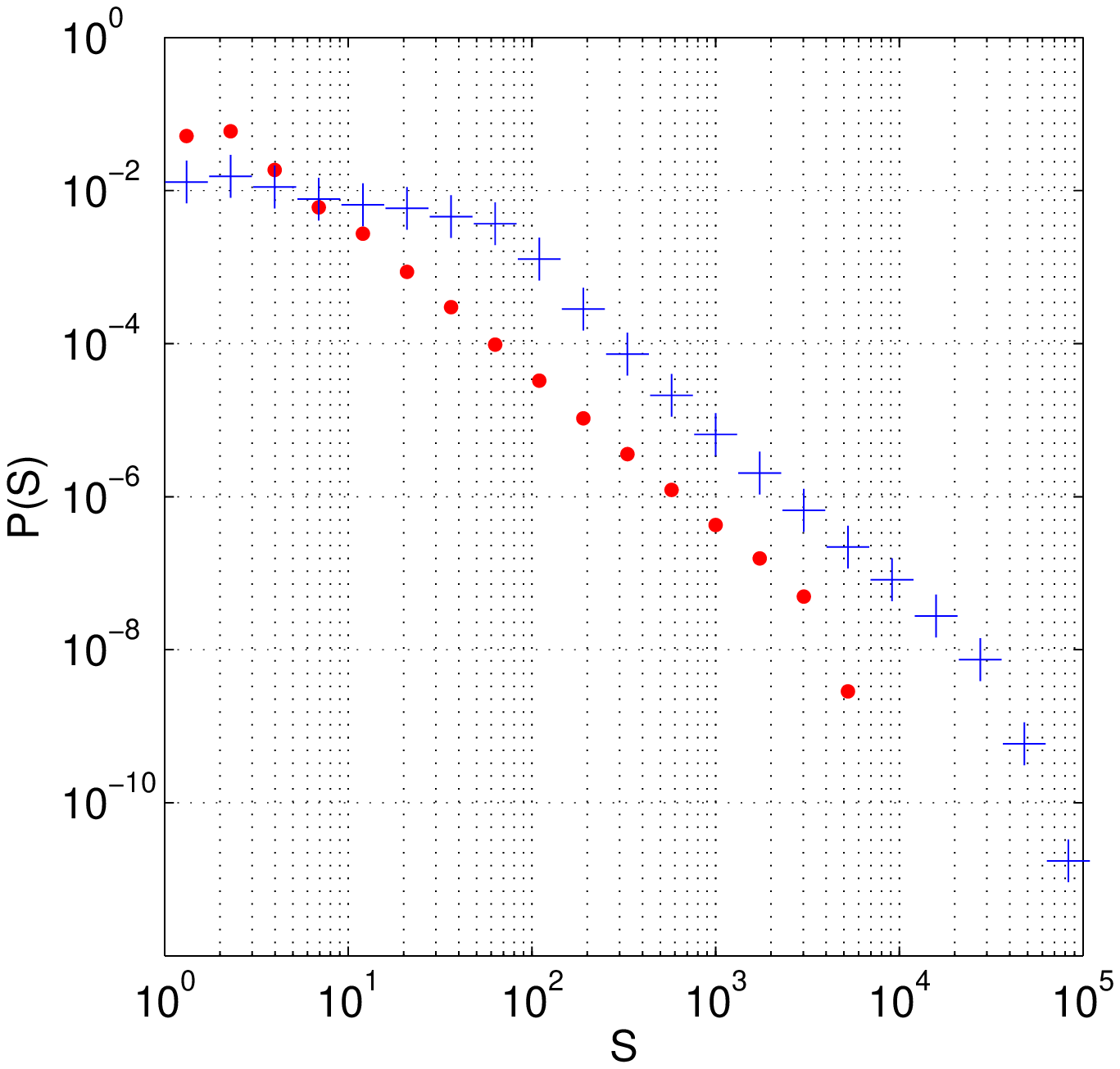}\hskip 3cm
\includegraphics[width=0.4 \textwidth]{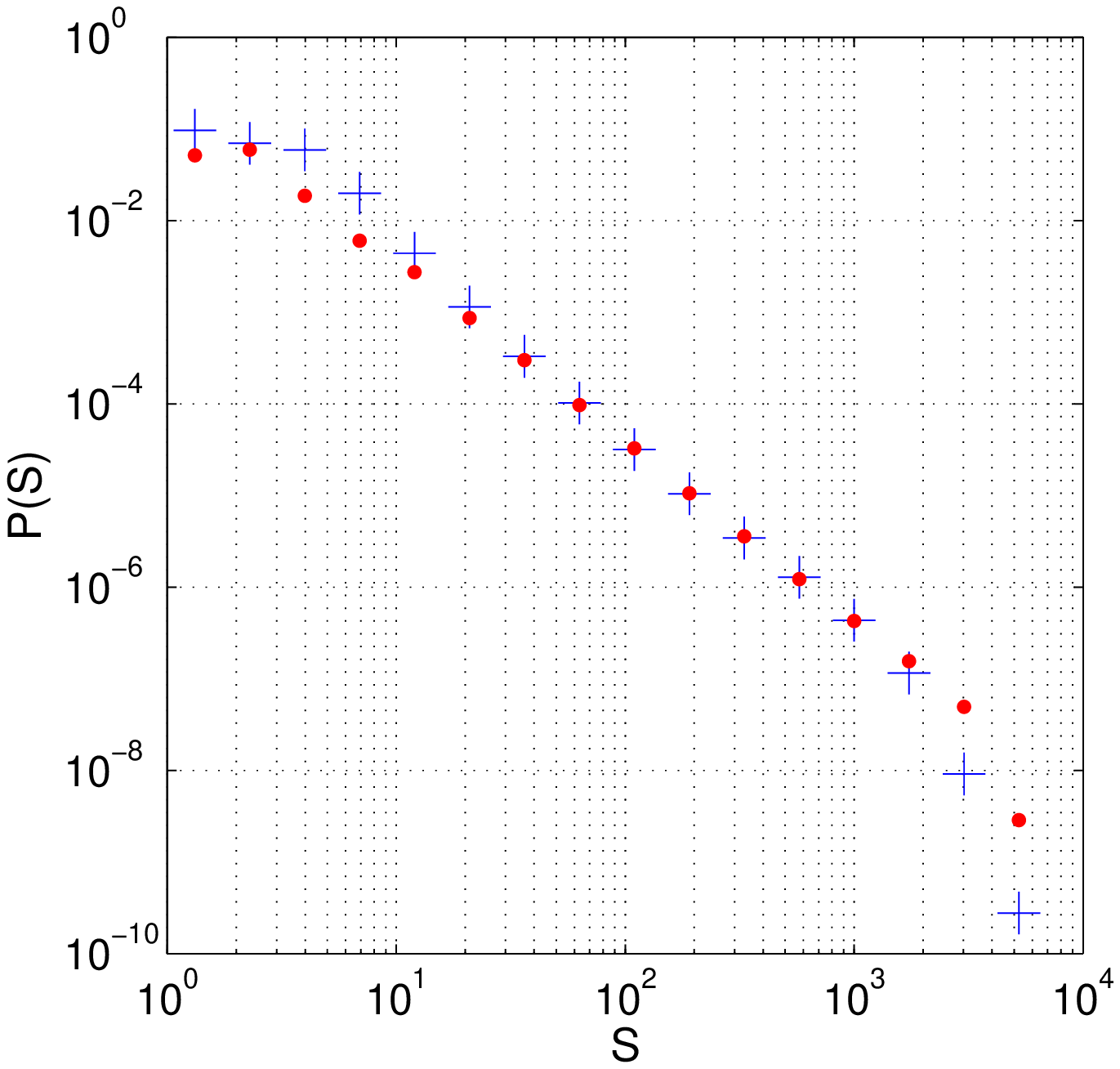}
\caption{Avalanche size normalized distributions for
$L_0=100$, $h=4$ ($\bullet$) and $L_0=400$, $h=16$ ($+$); (a-left)
probability densities; (b-right) as (a) with probability density for the
$h=16$ avalanche sizes rescaled $S\rightarrow S/16$.}
\end{figure}
This establishes a general property of avalanching systems  that
has been seen in several representative SOC
models\cite{corral,nickgrl,kinouchi,uritsky}, see also \cite{hwakardar,woodard}.

The above assumes the BTW case where avalanches relax instantaneously. One can develop this idea to introduce ``running sandpiles" (as extensively studied by, for example, \cite{hwakardar,corral,uritsky,woodard}) where redistribution is no longer instantaneous. Instead, wherever the critical gradient is exceeded locally, $f$ grains are moved in time $\delta t_{f}$. Thus there is a local redistribution rate per cell $h_f=f/\delta t_f$ which we can compare with the driving rate $h$. This introduces a new dimensionless parameter $(h \delta t_f)/(h_f \delta t)$ which modifies $h$ in the discussion above; in other words the SDIDT limit is approached for both $h \rightarrow 0$  and $\delta t_f/\delta t \rightarrow 0$ \cite{vesp,vesp2}. However, consistent with studies of running sandpiles \cite{hwakardar,corral,uritsky,woodard}, we can anticipate that avalanching phenomenology will persist for a range of finite $\delta t_f/\delta t$.

Depending on the
details, some SOC systems may show scaling in systems where the
drive is in fact highly variable. One could argue (see also \cite{vesp}) that such
robustness against fluctuations in the driver is necessary for SOC
to provide a `working model' in real physical systems where the
idealized SDIDT limit may not be realized.

\section{Control parameter for model ecosystem}
Finally, we apply the above framework  to simple models
for ecosystems. We consider a  large number of connected `meta- species' (or groups of species/variations occupying a given niche\cite{ecobook1})
with diverse sizes and rates of predation. Each meta- species is a
d.o.f. in the model. We insist that the
details are unimportant except that each of the $N$ meta- species,
 by acting as predator of one set of neighbors
in the food web and prey to another set, processes some dynamical
quantity, say, biomass or energy. The ecosystem then has a driving
rate, or rate of supply $H$ of biomass/energy  per unit volume at
the `bottom' of the web and a dissipation rate, or rate of
consumption $P$ of biomass by the top predators. We consider a
steady state on the average which includes secular change in these
parameters that is slow compared to the timescale for information
to propagate through the web.
For a given habitat, the abundance of species (i.e the relative
number of distinct meta- species) grows with the size of the
habitat. Although the details may vary, a good working
approximation for the `species- area relationship'\cite{rbook} is
a power law, so that the number of species $N$ in a habitat of
size $L_0$ is given by $N\sim (L_0^2)^\gamma$. A dimensionally
balanced expression in $D$ Euclidean dimensions is:
\begin{equation}
N\sim (L_0/L_c)^{D\gamma}
\end{equation}
with $\gamma >0$. The  length scale $L_c$
captures details of the sampling, as well as specifics of a given
habitat and terrain.
\begin{table}[t]
\begin{center}
\caption{$\Pi$ theorem applied to a simple model for an
ecosystem in a space with Euclidean dimension $D$. Interactions
between species processes a quantity (biomass, here) with
physical dimension $B$.}
\begin{tabular}{ccl}
\hline Variable        &dimension    & description  \\
\hline
$L_0$&$L$&system size\\
$L_c$&$L$&normalization length scale\\
$M_p$&$BT^{-1}$& top predator rate of consumption
\\
&&of biomass over system\\
$M_f$&$BT^{-1}L^{-D}$&rate of supply of biomass/unit volume\\
\hline
\end{tabular}
\label{tab1}
\end{center}
\end{table}
The relevant system variables are shown in Table 3. There are two dimensionless
groups:
\begin{equation}
\Pi_1=\frac{P}{H L_c^{D}}=R_B,\hskip 2pt
\Pi_2=\frac{L_0}{L_c}=f(N)=N^\frac{1}{D \gamma}
\end{equation}
thus we identify a control parameter $\Pi_1=R_B$ for the
simple ecosystem. To relate this to the abundance of species we
require some conservation property and to insist on steady state.
One possibility is to conserve some fraction of the biomass flux
propagated through the web so that for a steady state for the
system as a whole, the rate of supply of biomass is balanced by
the rate of removal by the top predators giving $L_0^DH \sim P$. In a system with losses, provided a fraction $\alpha$ of the
propagated quantity is on average passed from one d.o.f. or
species to the next, this expression is  $A\alpha^N L_0^DH
\sim P$; the factor $A$ also includes any recycling of the top
predator biomass to the bottom of the web.

We can however work
with any quantity which is transferred from one species to another
with some conservation. If we instead
consider $P$ and $H$ to refer to integrated energy consumption of
the top predator population and the energy taken up by the web per
unit volume (the productivity) respectively, conservation is then
the original `energetic- equivalence rule'\cite{wright}- that the
total energy flux of a population is invariant with respect to
body size.  The control parameter $R_B$ increases with a
measure of the rate at which biomass (or energy) is utilized by
the system as a whole ($P$),that is, is ultimately consumed by the top predator. Equivalently, it
increases with the biomass (or energy) rate of supply to the
system via the organisms at the bottom of the web, $H L_0$; these
both represent the rate at which biomass/energy is processed by
the ecosystem as a whole. We then have:
\begin{equation}
R_B\sim \left(\frac{L_0}{L_c}\right)^D\sim N^\frac{1}{\gamma}
\end{equation}
so we obtain that $R_B \sim N^{\beta_N}$ and quite generally that $\beta_N>0$.
The abundance of species simply increases  with  $R_B$ capturing the observation that diversity grows with
the global flux of energy/biomass, that is, productivity times
area\cite{wright,rbook}. This result holds even if the species-
area relationship is not a power law, it simply requires that the
number of species grows with habitat size; intriguingly, a non- power law $\beta_N$ suggests a
length scale dependence in the abundance of species that is intermittent.

 The power law
dependence implied by a power law species-area relationship
suggests that the dependence of $R_B$ on $N$ is rather nonlinear. A consequence is that, if we
consider slowly increasing this control parameter in a manner that
does not violate our assumption of a steady state, we would
expect, starting from an initial state of few species, to see a
sudden `explosion' in diversity at some critical value $R_L$.
 This will depend on the details through
the non universal parameter $L_c$; but since $L_c$ can be
determined through species- area abundance relationships, $R_L$
can in principle be determined.

We can consider
the analogy $R_B \equiv R_E$, that is, we identify $R_B$ as the
ecosystem's 'effective Reynolds number' which increases with the energy/biomass taken up by the system.  We have then shown that,
the disorder, or complexity of the ecosystem as expressed by the abundance of species increases with effective Reynolds number in the same sense as turbulence.
This analogy to turbulence may be instructive
in that there is some non universal value of the Reynolds number
at which a given system makes the transition  to turbulence. One can speculate that dynamical systems routes to
turbulence (specifically, the Ruelle-Takens or Feigenbaum
scenarios, e.g. \cite{george}) suggest a new approach to modelling
the onset of the diversity of species. Understanding ecosystems in the context of simple models for turbulence may also provide a basis for modelling the ``bursty" dynamics and scaling intrinsic to some ecosystems\cite{jorg}. Our approach to a
`generalized Reynolds number' outlined here potentially has
wider application: to living organisms and societal organizations,
insofar as they can be modelled\cite{ecobook1} as webs of many
interacting elements that process some dynamical quantity.
\ack
We thank A. Clarke, L. Demetrius, M. Freeman,  A. Neutel and R.
Williams for discussions,
 and the
STFC the EPSRC for support.

\end{document}